\documentclass[10pt]{article} 
\usepackage{graphicx} 
\usepackage{multirow}
\usepackage{caption}
\usepackage{subcaption}

\def\Title#1{
\begin{center}
	{\Large #1 } 
\end{center}
} 
\def\Author#1{
\begin{center}
	{ \sc #1} 
\end{center}
} 
\def\Address#1{
\begin{center}
	{ \it #1} 
\end{center}
}

\newcommand\pubblock{\rightline{
\begin{tabular}
	{l} Proceedings of the Second Annual LHCP\\
	\pubnumber\\
	\pubdate 
\end{tabular}
}}

\newenvironment{Abstract}{
\begin{quotation}
	\begin{center}
		\large ABSTRACT 
	\end{center}
	\bigskip 
	\begin{center}
		\begin{large}
			}{
		\end{large}
	\end{center}
\end{quotation}
}

\newenvironment{Presented}{
\begin{quotation}
	\begin{center}
		PRESENTED AT
	\end{center}
	\bigskip 
	\begin{center}
		\begin{large}
			}{
		\end{large}
	\end{center}
\end{quotation}
}

\def\beq{\begin{equation}}
\def\eeq#1{\label{#1}\end{equation}}
\def\eeqn{\end{equation}}

\def\beqa{\begin{eqnarray}}
\def\eeqa#1{\label{#1}\end{eqnarray}}
\def\eeqan{\end{eqnarray}}

\let\bar=\overbar

\def\Dslash{\not{\hbox{\kern-4pt $D$}}}
\def\dslash{\not{\hbox{\kern-2pt $\del$}}}

\def\msb{{\bar{\ssstyle M \kern -1pt S}}}

\newcommand{\pt}{\ensuremath{p_T}} 
\newcommand{\MET}{\ensuremath{E_\mathrm{T}^{miss}}} 
\newcommand{\HT}{\ensuremath{H_\mathrm{T}}} 
 
\newcommand{\GeV}{GeV} 
\newcommand{\LSP}{\ensuremath{\widetilde{\chi}_1^0}}
\newcommand{\njets}{\ensuremath{N_{\mathrm{jets}}}}
\mathchardef\mhyphen="2D
\newcommand{\nbjets}{\ensuremath{N_{\mathrm{b\mhyphen jets}}}}
\newcommand{\chargino}{\ensuremath{\widetilde{\chi}_{1}^{\pm}}}

\usepackage{color}

\newcommand\pubnumber{CMS-CR-2014/165}

\newcommand\pubdate{\today}

\def\affiliation{ On behalf of the CMS Collaboration. \\
Department of Physics \\
Universidad de Oviedo, 33006 Oviedo, Spain}

\begin{document}

\large 
\begin{titlepage}
	\pubblock
	
	\vfill \Title{Search for new physics using events with two same-sign isolated leptons in the final state in pp collisions at 8 TeV.} 
	\vfill
	
	\Author{ Santiago Folgueras } \Address{\affiliation} 
	\vfill 
	\begin{Abstract}

A search for new physics is performed based on events with jets and a pair of isolated, same-sign leptons. The results are obtained using a sample of proton-proton collision data 	collected by the CMS experiment at a centre-of-mass energy of 8 TeV at the LHC, corresponding to an integrated luminosity of 19.5 inverse femtobarns. In order to be sensitive to a wide variety of possible signals beyond the standard model, multiple search regions defined by the missing transverse energy, the hadronic energy, the number of jets and b-quark jets. No excess above the standard model background expectation is observed and constraints are set on a number of models for new physics.

	\end{Abstract}
	\vfill
	
	\begin{Presented}
		The Second Annual Conference\\
		on Large Hadron Collider Physics \\
		Columbia University, New York, U.S.A \\
		June 2-7, 2014 
	\end{Presented}
	\vfill 
\end{titlepage}
\def\thefootnote{\fnsymbol{footnote}} \setcounter{footnote}{0}

%
\normalsize 

\section{Introduction} \label{sec:intro} Events with same sign dilepton final states are very rare in the SM context, but they appear naturally in many different new physics scenarios. We describe searches for new physics with same-sign dileptons ($ee$, $\mu\mu$ and $e\mu$) and hadronic jets, with or without accompanying missing tranverse energy (\MET). Our choice of signature is driven by the following consideration. New physics signal with large cross sections are likely to be produced by strong interactions, and we thus expect significant hadronic activity in conjunction with the two leptons. Astrophysical evidence for dark matter, suggests that considering models with R-parity conservation, which provides an excellent dark matter candidate -- with a a stable lightest supersymmetric particle (LSP) that escapes detection. Therefore, a search for this signature involves sizable \MET due to undetected LSPs. 

	We consider several final states, characterized by the scalar sum (\HT) of the transverse momenta (\pt) of jets, \MET, the number of jets, and the number of jets identified as originating from b quarks (b-tagged jets). Having found no evidence of BSM contribution to the event counts, we set limits on a variety of SUSY scenarios by performing a counting experiment in each exclusive search region. What we present here is just a short summary of the analysis performed by CMS~\cite{CMS} as described in the original publication\cite{RA5}.

\section{Search strategy}\label{sec:eventsel} We require two isolated same-sign leptons ($e$ or $\mu$) with \pt $>$ 20 \GeV, consistent with originating from the same vertex. Events are collected using dilepton triggers and an extra veto on the third lepton is applied to suppress Drell-Yan production. The isolation of the leptons is computed with particle-flow information, and an event-by-event correction is made to account for the effect of the pileup. Additional cuts are applied on \MET\ , \HT\ , \njets\ and \nbjets\ to define the 24 exclusive search regions as defined in Table~\ref{tab:srs} that would be used in this search.

\begin{table}[h]
	\begin{center}
		\footnotesize	  
	\begin{tabular}{c|c|c|c|c}
		\hline \hline
		         \nbjets           & \MET  (\GeV)         & \njets   & \HT $\in [200,400]$~(\GeV) & $\HT >400$ (\GeV) \\ \hline
		\multirow{4}{*}{$=0$}      & \multirow{2}{*}{50--120} 	& 2--3      &    SR01      &    SR02     \\ 
		                           &                          	& ${\geq}4$ &    SR03      &    SR04     \\ \cline{2-5}
		                           & \multirow{2}{*}{${>}120$} 	& 2--3      &    SR05      &    SR06     \\ 
		                           &                         	& ${\geq}4$ &    SR07      &    SR08     \\ \hline
		\multirow{4}{*}{$=1$}      & \multirow{2}{*}{50--120} 	& 2--3      &    SR11      &    SR12     \\ 
		                           &                         	& ${\geq}4$ &    SR13      &    SR14     \\ \cline{2-5}
		                           & \multirow{2}{*}{${>}120$} 	& 2--3      &    SR15      &    SR16     \\ 
		                           &                         	& ${\geq}4$ &    SR17      &    SR18     \\ \hline
		\multirow{4}{*}{$\geq 2$}  & \multirow{2}{*}{50--120} 	& 2--3      &    SR21      &    SR22     \\ 
		                           &                         	& ${\geq}4$ &    SR23      &    SR24     \\ \cline{2-5}
		                           & \multirow{2}{*}{${>}120$} 	& 2--3      &    SR25      &    SR26     \\ 
		                           &                         	& ${\geq}4$ &    SR27      &    SR28     \\
                \hline \hline
		\end{tabular}
	\end{center}
	\caption{Definition of the signal regions uses the following numbering scheme:  the first digit in the name represents the requirement on the number of b-tagged jets for that search region.}
	\label{tab:srs}
\end{table}

\section{Background estimation}\label{sec:bkg} There are three main sources of SM background in this analysis, which are described below. More details on the methods used to estimate these backgrounds can be found in the original documentation.\cite{RA5}

\begin{itemize}
	\item \textbf{Background with one or two fake leptons:} Include processes such as semi-leptonic $\mathrm{t\bar{t}}$ or $W+jets$ where one of the leptons comes from a heavy-flavor decay, misindentified hadrons, muons from light-meson decay in flight, or electrons from unidentified photon conversions. We estimate this background starting from measuring the probability of a lepton being fake or prompt using a QCD or Z enriched sample respectively. We then apply those probabilities to events passing the full kinematic selection but in which one or two of the leptons fail the isolation requirements. About 40-50\% of the total background is due to this processes and we assign a 50\% systematic uncertainty to account for the lack of estatistics in the control sample as well as the little knowledge we about about the control sample composition. 
	\item \textbf{Events with charge mis-identification:}  These are events with opposite-sign isolated leptons where one of the leptons (typically an electron) and its charge is misreconstructed due to sever bremsstrahlung in the tracker materia (this effect is negligible for muons). We estimate this background by selecting opposite-sign $ee$ or $e\mu$ events passing the full kinematic selection, weighted by the probability of electron charge misassignment. This probability is measured in a $Z\rightarrow e e$ sample in data by simply calculating the ratio between same-sign and opposite-sign events in such sample and it validated in MC, this probability is of the order 0.02 (0.2)\% for electrons in the barrel(endcap). This source of background only only accounts for the 5-10\% of the total background. A 20\% systematic uncertainty on this background is considered to account for the \pt\  dependence of the probability. 
	\item \textbf{Rare SM processes:} These include SM processes that yield two same-sign prompt leptons, including $\mathrm{t\bar{t}W}$, $\mathrm{t\bar{t}Z}$, $\mathrm{W}^\pm\mathrm{W}^\pm$ among others. These processes constitutes about 30-40\% of the total background. All these backgrounds are obtained from Monte Carlo simulation and we assign a 50\% systematic uncertainty to this background sources as we have very little knowledge on the cross-sections. 
\end{itemize}

\section{Results}\label{sec:results}
\begin{figure}[htb]
\centering
\includegraphics[width=0.325\textwidth]{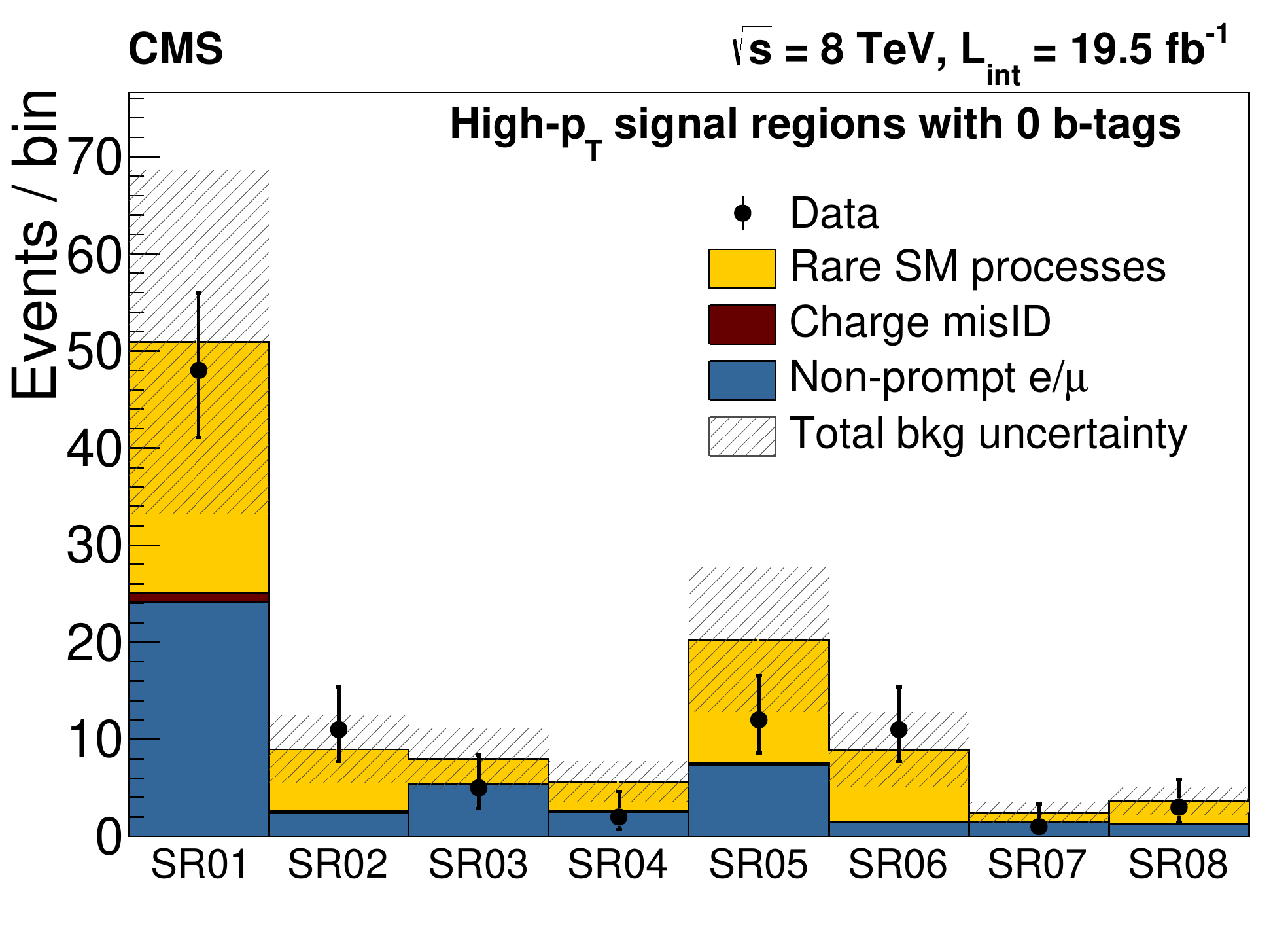}
\includegraphics[width=0.325\textwidth]{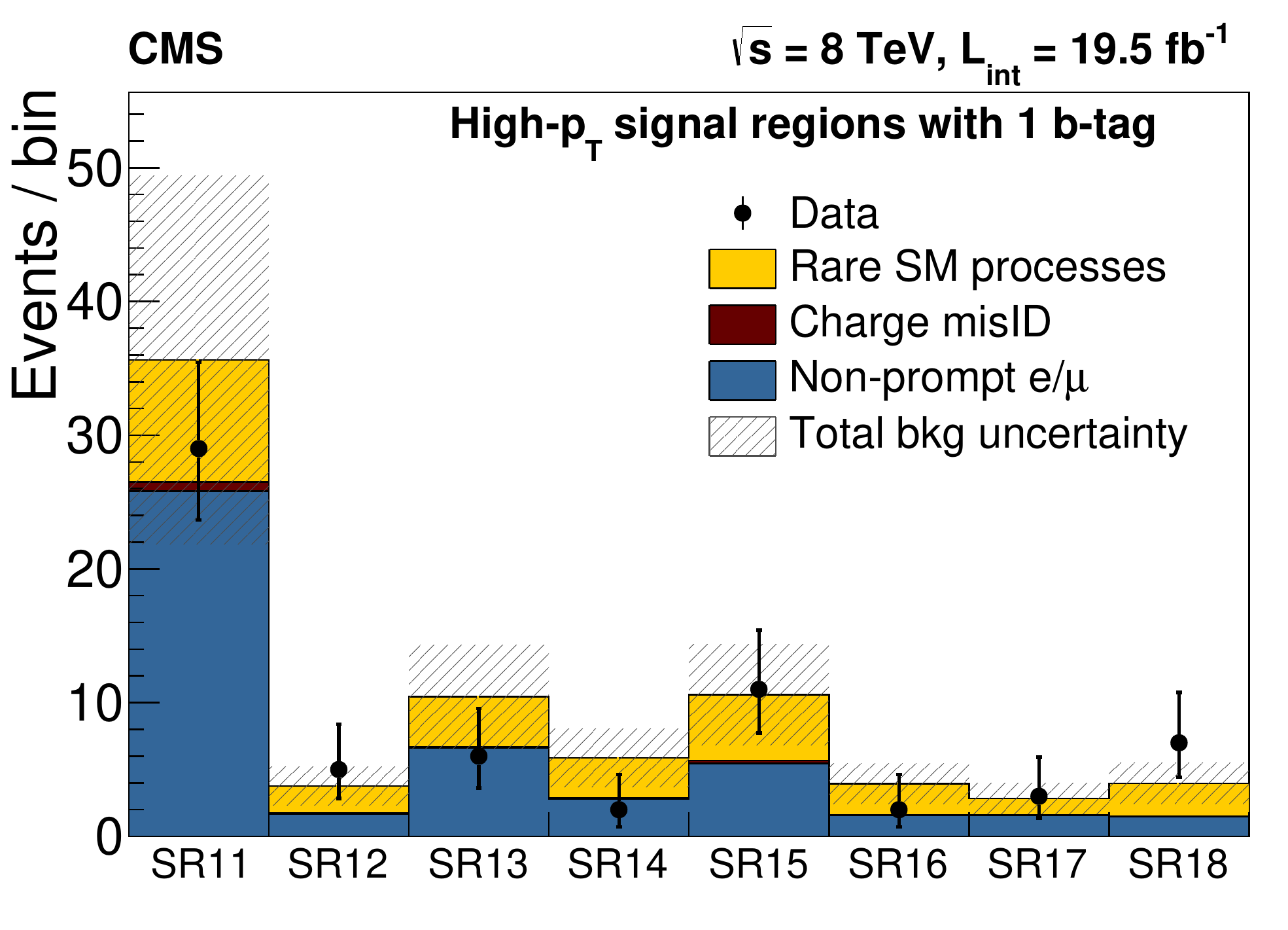}
\includegraphics[width=0.325\textwidth]{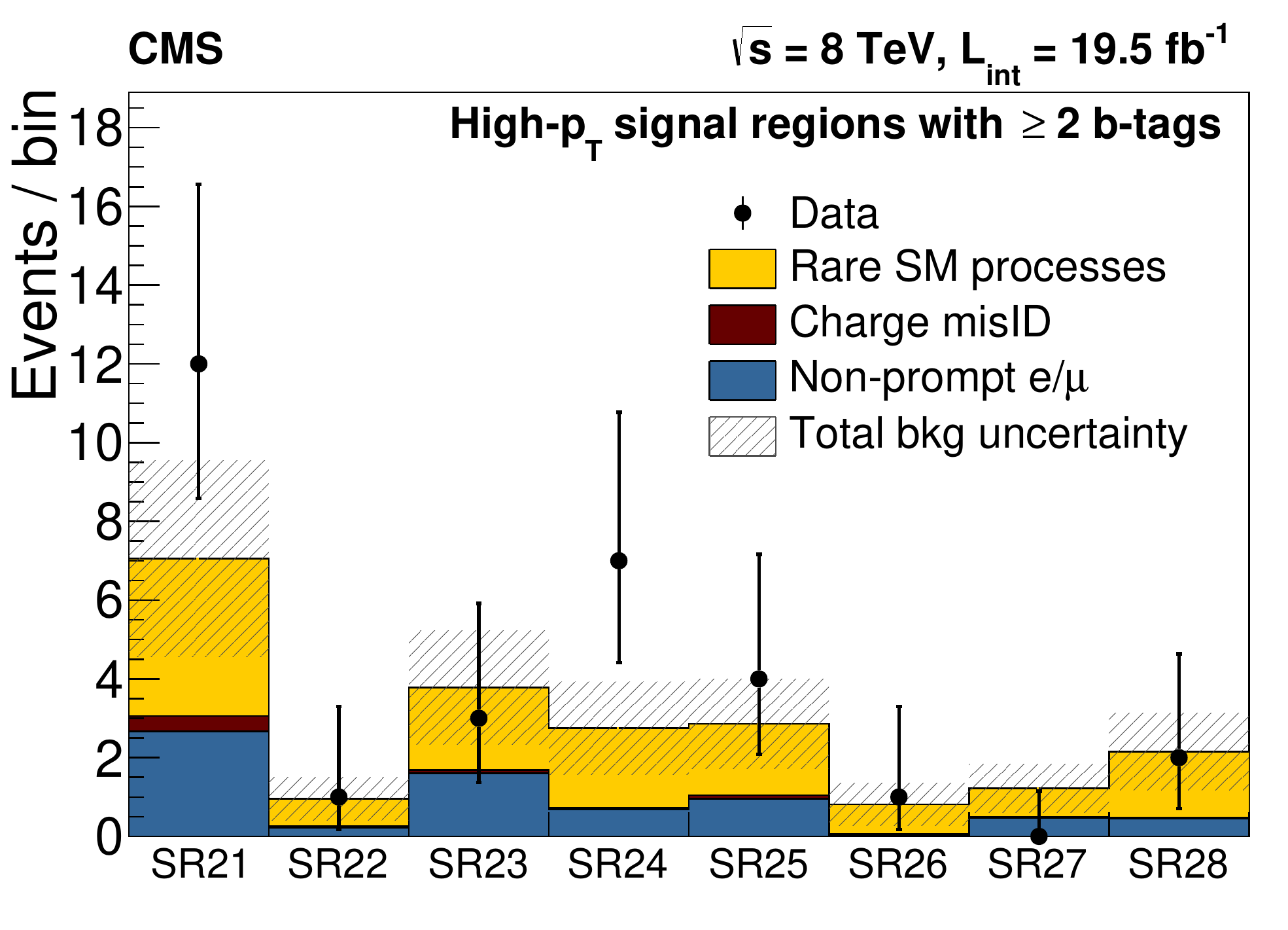}
\caption{Summary plots showing the predicted background from each source and observed event yields for all the different SRs. The shaded region represents the total background uncertainty.}
\label{fig:SRs}
\end{figure}

The observations in each of the final signal regions are presented in Fig. \ref{fig:SRs} along with the corresponding SM background prediction. The contributions of rare SM proccesses and non-prompt leptons vary among the signal regions between 40\% and 60\%, while the charge misidentification background is almost negligible for all signal regions. The observations are consistent with the background expectations within their uncertainties. 

\section{Interpretations}\label{sec:inter}
Tthe results of the search are used to derive limits on the parameters of various models of new physics. A combination of the most sensitive SRs are used to interpret the results in different scenarios. We only show here a glimpse of the various scenarios that we consider, more scenarios are covered in the original documentation~\cite{RA5}.

\begin{figure}[htb]
\centering
        \begin{subfigure}[b]{0.3\textwidth}
                \includegraphics[width=\textwidth]{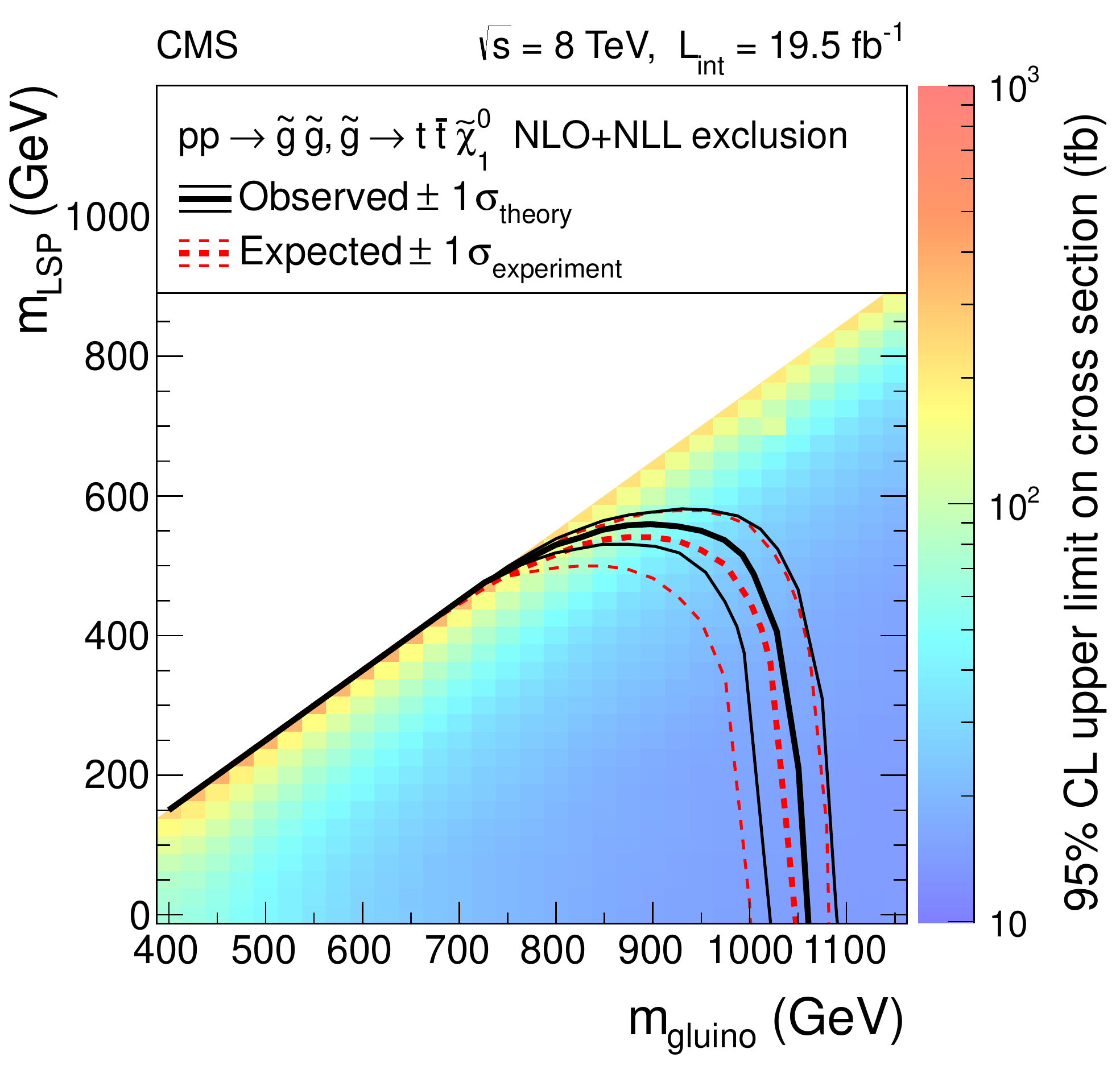}
				\caption{}
                \label{fig:A1}
        \end{subfigure}%
        \begin{subfigure}[b]{0.3\textwidth}
                \includegraphics[width=\textwidth]{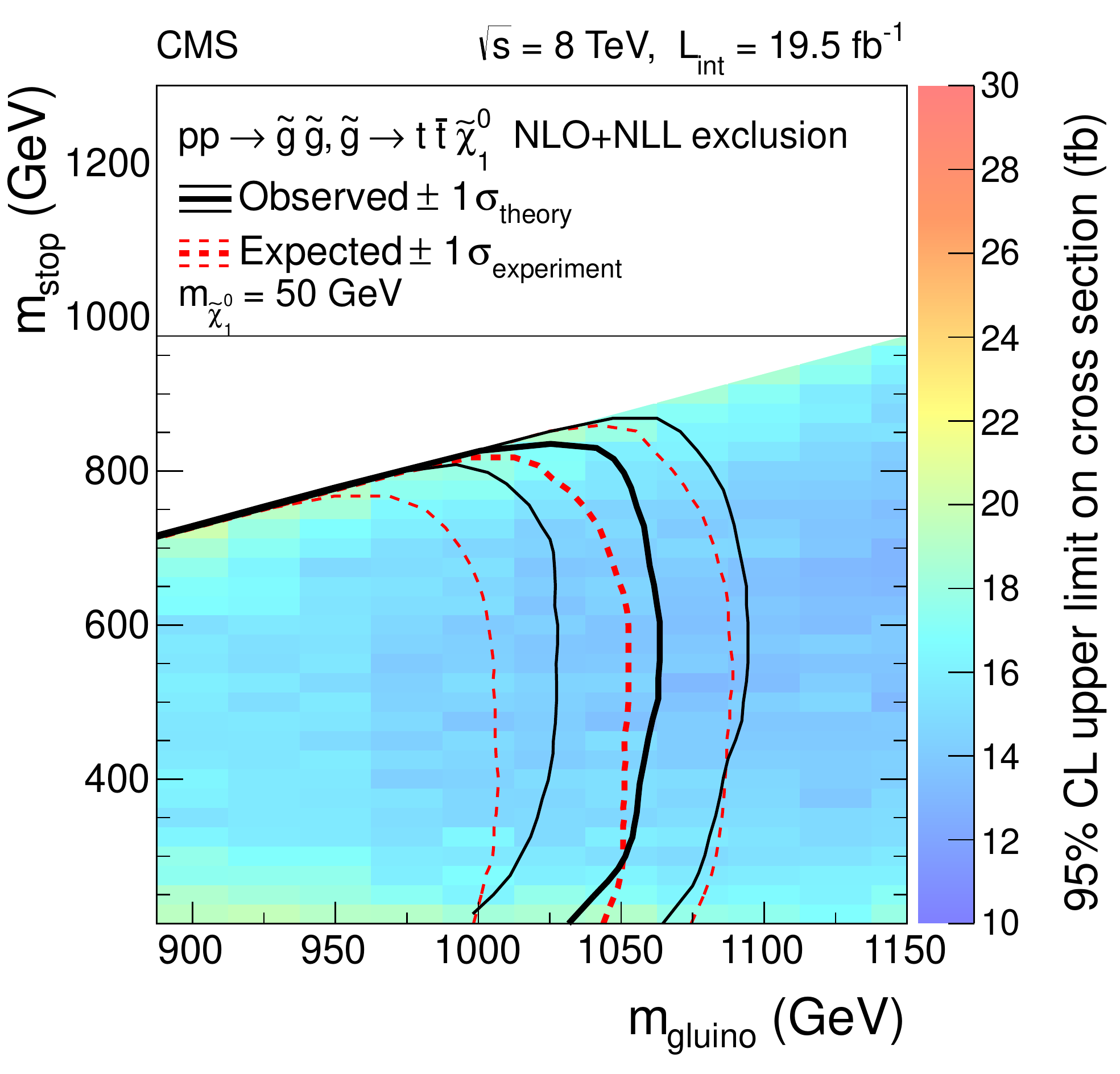}
				\caption{}
                \label{fig:A2}
        \end{subfigure}\\
        \begin{subfigure}[b]{0.3\textwidth}
                \includegraphics[width=\textwidth]{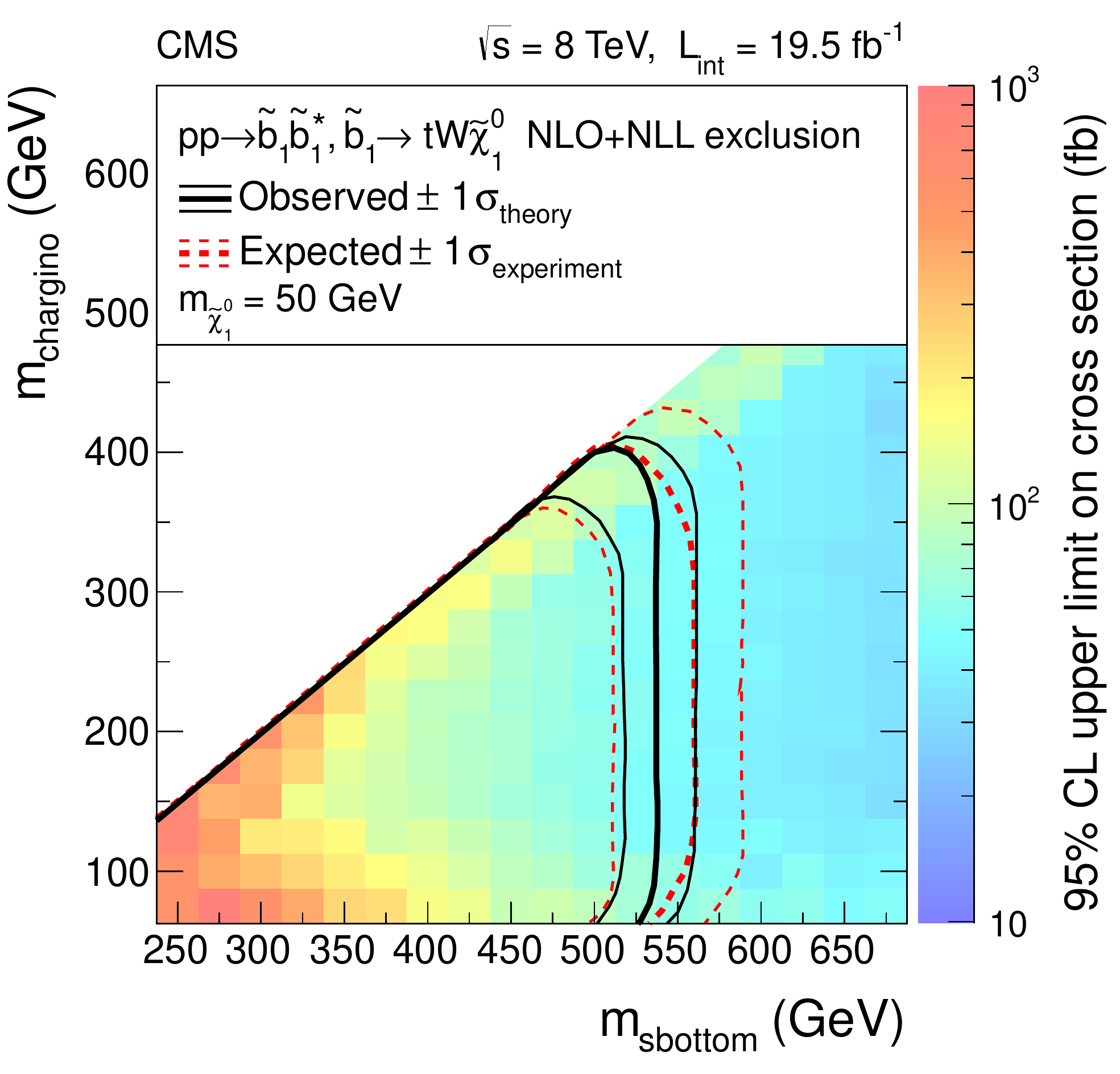}
				\caption{}				
                \label{fig:B1a}
        \end{subfigure}%
        \begin{subfigure}[b]{0.3\textwidth}
                \includegraphics[width=\textwidth]{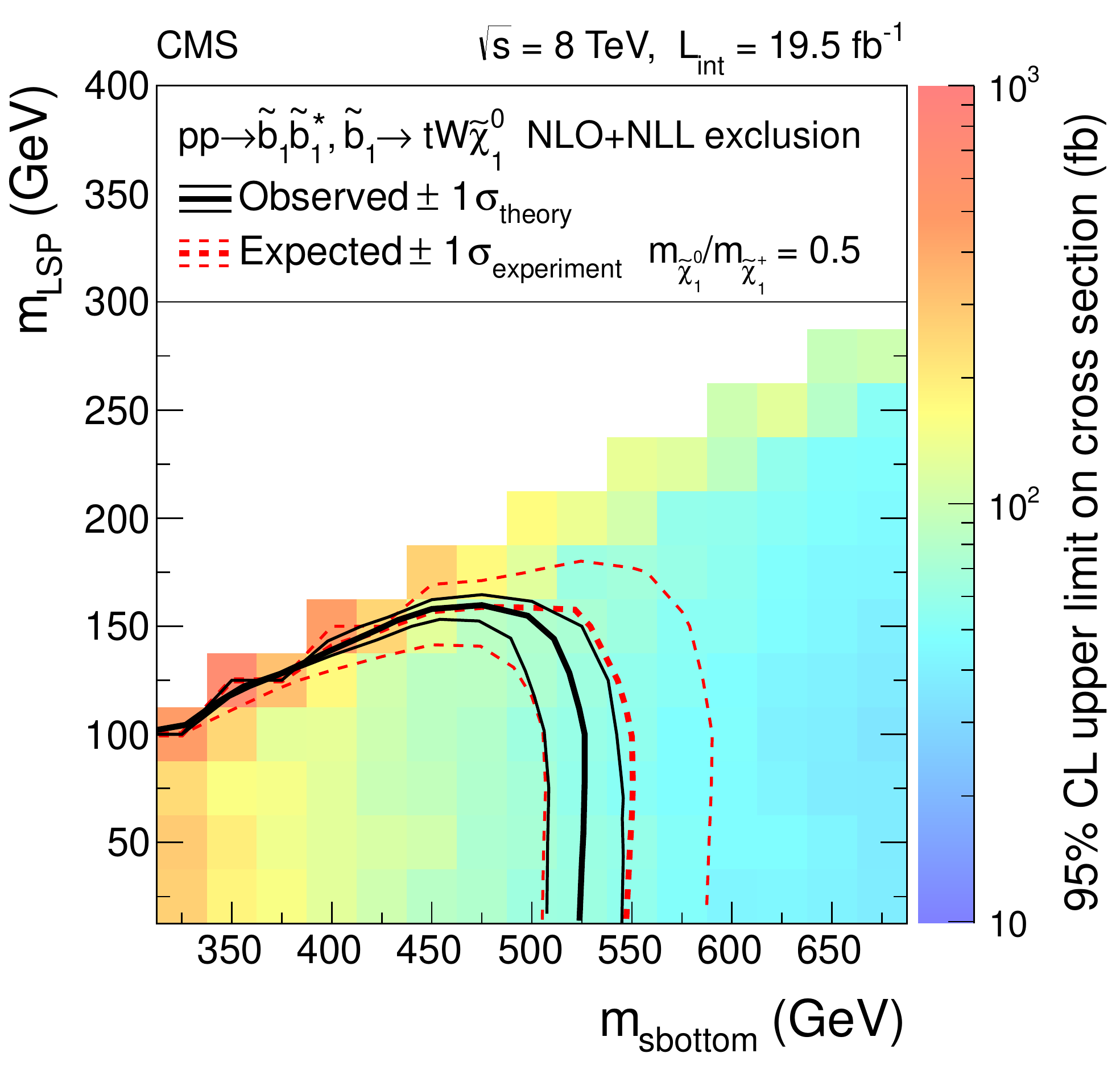}
				\caption{}
                \label{fig:B1b}
        \end{subfigure}%

\caption{Exclusion regions at 95\% CL in the planes of (top-left) m(\LSP) versus m(\ensuremath{\widetilde{g}}), (top-right) m(\ensuremath{\widetilde{t}_1}) versus
m(\ensuremath{\widetilde{g}}), (bottom-left) m(\chargino) versus m($\widetilde{b}_1$), and (bottom-right) m(\LSP) versus m($\widetilde{b}_1$). The excluded regions are those within the kinematic boundaries and to the left of the curves. The effects of the theoretical uncertainties in the NLO+NLL calculations of the production cross sections are indicated by the thin black curves; the expected limits and their $\pm$1 standard-deviation variations are shown by the dashed red curves.}
\label{fig:model}
\end{figure}

For the interpretation of the results we consider here a model with gluino pair production in the tttt\LSP\LSP\ final state where \LSP\ is the lightest neutralino. The gluino either undergoes a three-body decay mediated by an off-shell top squark (Fig.~\ref{fig:A1}) or a two-body decay to a top quark and a top anti-squark, that will decay into a top quark (Fig.~\ref{fig:A2}). Other scenario that we explore is the sbottom pair production, followed by one of the most likely decays of the sbottom, $\widetilde{b}_1\ \rightarrow\ t\chargino\ $ with $\chargino \rightarrow W\LSP\ $. We consider two cases in this decay mode. We either set the \LSP\ mass to 50 GeV and present the limits in the ($m_{\chargino}$, $m_{\widetilde{b}_1}$) plane (Fig.~\ref{fig:B1a}), or consider the ($m_{\LSP}$,$m_{\widetilde{b}_1}$) plane (Fig.~\ref{fig:B1b}) with the mass of the chargino to be twice the mass of the \LSP, so the top quark and W boson are on shell. Some other scenarios can be found on the original document \cite{RA5}. The exclusion limits in these scenarios can be found in Figure~\ref{fig:model}.

\section{Conclusions}\label{sec:conc} 
We have presented results of a search for new physics with events with same-sign dileptons using the CMS detector at the LHC. No significant deviations from the standard model expectations are observed. The results are used to set exclusion limits into several SUSY models. We are able to probe gluino (sbottom) masses up to 1050 (500) GeV.



\begin{thebibliography}{99}
	\bibitem{CMS}
	  CMS Collaboration,
	  JINST {\bf 3} (2008) S08004.
	
	\bibitem{RA5} 
	  CMS Collaboration,
 	 JHEP {\bf 1401}, 163 (2014)
 	 [arXiv:1311.6736, arXiv:1311.6736 [hep-ex]].


\end{thebibliography}
\end{document}